\documentstyle[a4,psfig]{article}
\begin{document}

\renewcommand{\baselinestretch}{1.5}

\begin{center}
{\Large{\textbf{Action Correlations and Random Matrix Theory \vspace{10mm}}}}\\
{\large{ Uzy Smilansky and Basile Verdene \vspace{5mm}}}\\
{Department of Physics of Complex Systems\\ Weizmann Institute of
Science, Rehovot 76100, Israel \vspace{5mm}\\email:
uzy.smilansky@weizmann.ac.il, b.verdene@weizmann.ac.il
 \\[2mm]}
\end{center}

\date{\today}


\begin{abstract}
 The correlations in the spectra of quantum systems are intimately related to correlations which are of
genuine classical origin, and which appear in the spectra of actions of the classical periodic orbits of the
corresponding classical systems. We review this duality and the semiclassical theory which brings it about.
The conjecture that the {\it quantum} spectral statistics are described in terms of Random Matrix
Theory, leads  to the  proposition that the {\it classical} two-point correlation function is given also in terms
of a universal function. We study in detail the spectrum of actions of the Baker map, and use it to illustrate the
steps needed to reveal the classical correlations, their origin and their relation to symbolic dynamics.

\end{abstract}
\section{Introduction}
\label{introduction}

The interest in the correlation between classical periodic orbits, and
in particular, in the spectrum of their actions, emerges from the
attempts to provide a semiclassical proof of the universality  of spectral correlations of
quantum chaotic systems (``the BGS conjecture" \cite {bgs}).  Action correlations
were first discussed  by Argaman {\it et. al.} were the universality of action correlations, and their
relation to Random Matrix Theory (RMT) were studied for a few chaotic systems
\cite{argamandittes}. Various aspects of the subject were investigated later
\cite {dittesdoron,aurich,tanner,cohenprimack,sano,primack,smil2}.
This culminated  recently in the work of Sieber and Richter \cite
{SR2001}, who identified pairs of correlated trajectories,
whose contribution to the spectral form factor for systems with time
reversal symmetry is identical to the next to leading order term in
the formfactor predicted by RMT (see also \cite {Braun2002,Berko2002}).

The purpose of this paper is twofold: first, to review the semiclassical context  where
action correlations and their expected universal features arise in a natural way \cite
{primack,smil2}. This will be done in the main body of the present section, where also the
connection with RMT will be made explicit.
Second, to test the general semiclassical arguments on action correlations for a paradigm
chaotic dynamical system - the Baker map. This system was investigated previously by
a number of groups, \cite {argamandittes,dittesdoron,tanner,sano}, who demonstrated
numerically the existence of  the expected correlations.  Here, we develop  another approach for the
analysis of the action spectrum, where we try to systematically asses the way the periodic orbits
and their actions can be partitioned to families which are dynamically related. One aspect
of this approach is studied by casting the problem as an Ising model (in one
dimension and with a long range, yet exponentially decaying interaction). Moreover, we
show several features of the action correlations which escaped the attention of
previous works, and in particular, we  analyze  the correlations in terms of
the symbolic dynamics. We hope that this insight will pave the way to a more complete
understanding of the universal features of action correlations in general.

\subsection{Action correlations and the semiclassical theory of spectral statistics}
Consider a finite, two dimensional domain ${\cal A}$ and an area preserving map
\begin{equation}
 {\cal M} :
 \gamma'= {\cal M}(\gamma)\ \ ; \ \ \gamma', \gamma\in {\cal A}.
\end{equation} Any area preserving map
 can be expressed implicitly through a generating function \cite {gold}.  Let
$\gamma=(x,y) \ , \  \gamma'=(x',y')$. The generating function $F_1(x',x)$ defines the map
through the relations
\begin{eqnarray}
\label{equ:mapping}
y= - \frac {\partial F_1(x',x)}{\partial x}  \ \ ; \ \
y'= \ \  \frac {\partial F_1(x',x)}{\partial x'}
\end{eqnarray}
where $x'$ is to be expressed in terms of $(x,y)$ by solving the first equation,
and $y'$ is given explicitly using the second equation. Twist maps are  maps for which
the first equation above has a unique solution for any $(x,y) = \gamma \in {\cal A}$.
 From now on we shall assume  ${\cal M}$ to be a {\it hyperbolic} twist map.

Periodic points of period $n$ are solutions of the equation $\gamma= {\cal M}^n
(\gamma)$.  The corresponding $n$-periodic
 orbit is obtained by iterating the map:
\[\gamma_0^{(a)},\gamma_1^{(a)}={\cal M}(\gamma_0^{(a)}), \cdots ,
\gamma_{n-1}^{(a)}={\cal M}^{n-1}(\gamma_0^{(a)}).\]
 The {\it action} assigned to a periodic orbit is
\begin{equation}
f_a =\sum_{j=1}^n F_1(x_{j}^{(a)},x_{j-1}^{(a)})\ \  \ \ ;\ \ \  \ x_0^{(a)}=x_n^{(a)}
\end{equation}
The number of $n$-periodic orbits, $N_n$ increases exponentially with $n$. The object of
our investigations is the set of actions $ \{f_a\}_{a=1}^{N_n}$,
in the limit $n\rightarrow \infty$.

Under quite general conditions one can show that the mean action $\left<
f\right>_n$ and the variance var$f(n)$ are both proportional to $n$.
Moreover, the actions of $n$-periodic orbits are bounded within an
interval which grows $\it algebraically$ with $n$. However, since their number
grows {\it exponentially} with $n$, one expects exponentially small spacings
between successive actions. We shall explain now why the pair
correlations of action are expected to be universal, and determined by RMT. To this
end we should address the quantum analogue of  the map, in the semiclassical limit.

 The quantum analogue of the classical map, is a unitary evolution operator
$U_N$ which acts on a $N$ dimensional Hilbert space, with
\begin{equation}
\label{eq:nnn}
 N=\left[ ||{\cal A}||\over {2\pi
\hbar}\right ] ,
\end{equation}
 where  $\left [
\cdot \right ]$ stands for the integer value and $||{\cal A}||$ is the area of ${\cal A}$.

The spectrum of $U_N$ consists of $N$ unimodular complex numbers ${\rm
e}^{i\theta_l(N)}, \theta_l(N)\in [0,2\pi)$. Ample numerical evidence supports the
conjecture that the  spectral statistics of $U_N$, is well reproduced by the predictions
of random matrix theory (RMT).   The quantum spectral density is denoted by
\begin{equation}
\label{eq:qmdens}
d_{qm}(\theta;N) \ = \sum_{l=1}^N \delta (\theta- \theta_l(N)).
\end {equation}

In the semiclassical approximation,
\begin{equation}
\langle  x'|U_N|x\rangle =\left ({1 \over2\pi \hbar i}\right ) ^{1\over 2}
\left ({ \partial^2 F_1(x',x) \over \partial x'
\partial x}\right ) ^{1\over 2}
{\rm e} ^{ i F_1(x',x) /\hbar} \ .
\label {sclsmatrix}
\end{equation}
Here,  $x$ is  a discrete index which labels the eigenstates of the position
operator and  $F_1(x',x)$ is the classical generating function.  A detailed discussion
of the semiclassical evolution operator and its properties can be found in
\cite{smil1}.

The semiclassical approximation (\ref {sclsmatrix}) yields
\begin{equation}
\label{eq:duality}
 \int_0^{2\pi} {\rm d}\theta \ {\rm e}^{i n \theta } d_{qm}(\theta;N)\nonumber
  =  {\rm tr}
U_N^n \approx \
  \sum_{ a\in {\cal P}_n } A^{(n)}_a {\rm e}^ {i f_a /\hbar + i \frac{\pi}{4}r_a \mu_a  }
 \nonumber
\end{equation}
where ${\cal P}_n$ is the set of  $n$-periodic orbits. $A^{(n)}_a$ are the stability
amplitudes
\begin{equation}
A^{(n)}_a = { n \over r_a |\det
              (I-T^{r_a}_a)|^{{1\over 2}} } \ .
\label{trsemicl}
\end{equation}
 Here, $r_a$ stands for the repetition number if the periodic orbit is a repetition of a primitive
 $n_a$-periodic orbit $(n_a=\frac{n}{r_a})$.  $T_a$ is  the monodromy matrix and $\mu_a$ is the
 Maslov index for the primitive orbit.

 At this point we introduce the {\it classical}  density of actions of $n$-periodic
orbits,
\begin{equation}
\label{eq:defdcl}
d_{cl}(f;n) \ = \sum_ { a\in {\cal P}_n} A^{(n)}_a {\rm e}^{ i  \frac{\pi}{4}r_a \mu_a }
 \delta (f- f_a(n)),
\end{equation}
and we obtain
\begin{equation}
\label{eq:dual2}
{\rm tr}
U_N^n \approx
  \int {\rm d}f \ {\rm e}^{i f /\hbar}\  d_{cl}(f; n) \ .
\end{equation}
Using (\ref{eq:nnn}), and measuring the actions in units of the phase space area
$s=f/||{\cal A}||$, we get the relation between the quantum and the classical densities:
\begin{equation}
\label{eq:dualdual}
\left .  \int_0^{2\pi} {\rm d}\theta \ {\rm e}^{i \nu \theta } d_{qm}(\theta;N)  \right
|_{\nu=n}
\approx
\left .  \int {\rm d}s \ {\rm e}^{i k s}\  d_{cl}(s; n) \right |_{k=2\pi N}\ .
\end{equation}
This equation expresses the quantum - classical duality. It relates two densities  which are
very different: The quantum density gives a unit weight to all the eigenphase on the unit
circle, while in the classical density, the actions are weighted by the stability amplitudes,
and are assigned with a ``charge" $\pm 1,\ \pm i$ depending on the Maslov index.
The duality relation is expressed via a Fourier transform which involves both the variables
and the parameters which specify the quantum and the classical distributions.    On  the
quantum side, $N$ is a {\it parameter} which specifies the value of $\hbar$, while $n$ is the
value of the variable conjugate to $\theta$. On the classical side, $n$ is a {\it parameter}
which specifies the period of the ensemble of orbits under consideration, while $N$ determines
the value of $k$ - the variable conjugate to $s$. In the sequel we shall always reserve the
first position to the spectral variable or its conjugate, while the second position is
reserved to the parameter which specifies the system.

We shall focus our attention on the spectral formfactors, which are the Fourier transforms of the
pair correlation functions. The quantum and the classical formfactors are given explicitly as
 \begin{equation}
\label{eq:qmff}
K_{qm}(\nu;N)=\frac{1}{N}\sum_{l,l'=1}^N {\rm e}^{i \nu (\theta_{l}-\theta_{l'} )}\  =
\frac{1}{N}\sum_{l,l'=1}^N {\rm e}^{ 2\pi i \tau (\theta_{l}-\theta_{l'} ) \bar d_N}\  \ ,
\end{equation}
where, $\tau =\frac{\nu}{N}$ and $\bar d_N = \frac{N} {2\pi}$ is the mean density and $\nu$ must be an integer
since $d_{qm}$ is a periodic function. The classical formfactor is,
 \begin{equation}
\label{eq:clff}
K_{cl}(k;n)=
  \sum_{ a,a'=1}^{N_n} A^{(n)}_a A^{(n)}_{a'}{\rm e}^ {i \frac{\pi}{4}(r_a \mu_a -
r_{a'} \mu_{a'})  }  {\rm e}^ {ik (s_{a}-s _{a'}  ) } \ ,
\end{equation}
 It follows from
(\ref{eq:dualdual}) that the quantum and the classical 2-point formfactors are related by
\begin{equation}
\label{eq:kkduality}
K_{qm}(\nu=n;N) \approx  \frac{1}{N} K_{cl}(k=2 \pi N;n) \ .
\end{equation}
Hence, the quantum spectral correlations are reflected in the
classical spectrum, and {\it vice versa}. This equation expresses the important semiclassical
result that the quantum formfactor is obtained from the classical one by interrogating the
spectrum of action differences  on the scale of $N^{-1} \sim  \hbar$.

Equation (\ref {eq:kkduality}) has to be understood in the sense of distributions,
since the formfactors, as defined above, are the Fourier transforms of a sum of
$\delta$ functions. Indeed, the formfactors computed  for a given system,
fluctuate, and do not converge to a limit when $N\rightarrow \infty$. The appropriate
way to overcome this difficulty in the present context is to apply a smoothing
procedure, which enables the extraction of  well defined  limit distributions.

Starting with the quantum formfactor, we adopt {\it spectral averaging} which is based
on the assumption of ``spectral ergodicity". We order the spectrum so that $
\theta_l\le\theta_{l+1}$, and partition it to
$N_g$ subsets $\sigma_{g}$, each consisting of $\hat N = \frac {N}{N_g}$  subsequent
phases. Neglecting correlations between phases in different subsets, we rewrite
(\ref{eq:qmff}) as
 \begin{eqnarray}
\label{eq:qmffsmoo}
K_{qm}(\tau;N)& \approx&\frac{1}{N_g}\sum _{ g=1}^{N_g} \frac{1}{\hat N }\ \sum_{l,l' \in
\sigma_{ g}} {\rm e}^{i 2\pi \tau (\theta_{l}-\theta_{l'} )\bar d_N}\nonumber \\
&=&
\frac{1}{N_g}\sum_{ g=1}^{N_g} K_{qm}^{(g)}(\tau;\hat N)  \ .
\end{eqnarray}
$K_{qm}^{(g)}(\tau;\hat N) $ is the formfactor for the
spectrum obtained by multiplying all $\theta_l \in \sigma_{g}$ by
$\frac{N}{\hat N}$ so that they cover  uniformly the entire circle. Taking now the limit
$N\rightarrow \infty$ at a constant $\tau$, with $N_g \approx \hat N  \approx \sqrt N$,
it is expected that $K_{qm}(\tau;N)$, which is expressed now as an average over the ensemble of the partial
formfactors, converges to a limit distribution
$K_{qm}(\tau)$ which reproduces the prediction of RMT for the appropriate ensemble. This procedure is
justified by the fact that in the quantum spectrum,
 the {\it correlation range} is of the order of a mean
spectral spacing. Hence, the elimination of the correlations between different sets in
(\ref{eq:qmffsmoo}) introduces a small error, which vanishes in the large $N$ limit.

A different smoothing procedure is required for the discussion of the classical
spectrum. As will be shown in the sequel, the classical correlation length is of order
of the {\it inverse} of ${n}$ which exceeds by far the mean spacing which is exponentially
small in  $n$. A spectral smoothing should therefore  rely on a different partitioning of the
period orbits, such that the relevant correlations are preserved within a subset, while
members of different subsets are statistically uncorrelated. One of the main
problems in dealing with the classical correlations is the proper definition of
such subsets  \cite {cohenprimack}. The smoothing of the classical spectra is by far
more important since the number of actions increases exponentially with $n$.

If we denote the results of the quantum and the classical smoothing procedures by
$\langle  \cdot \rangle$, we obtain from  (\ref{eq:kkduality}) the relation which is
basic to the present approach,
\begin{equation}
\label{eq:basiduality}
\langle K_{qm}(\tau;N)\rangle  \approx   \frac{1}{N}\langle
K_{cl}(2\pi\frac{n}{\tau};n)\rangle
\ \ \ ;
\ \ \
\tau =\frac{n}{N} \ .
\end{equation}
  Assuming that  $\langle K_{qm}(n;N)\rangle$ follows the RMT  predictions, namely,
$\langle K_{qm}(n;N)\rangle = K_{RMT}(\tau=n/N)$, we find that the classical spectra of
chaotic systems must display universal pair correlations which can be derived from RMT
using the relation (\ref {eq:basiduality}). In particular, since for large $N$, $\langle
K_{qm}(\tau;N)\rangle$ is a function of $\tau$ only, we derive two immediate predictions
for  $\langle K_{cl}(2\pi\frac{n}{\tau};n)\rangle $ :

\noindent {\it (i)} $\langle K_{cl}(2\pi\frac{n}{\tau};n)\rangle $ is proportional to
$n$, since the  {\it rhs} of (\ref{eq:basiduality})  must depend on $N$ only through the ratio
$\frac{n}{N}$.  It is convenient to define
\begin{equation}
\label{eq:kcal}
{\cal K}_n(k) = \frac{1}{n} \langle K_{cl}(k;n)\rangle \ ,
\end{equation}
where the proportionality of  $\langle K_{cl}(2\pi\frac{n}{\tau};n)\rangle $ to $n$ is made explicit.

\noindent {\it (ii)}   ${\cal K}_n(k)$  depends on $k$ only through the
scaled variable $\frac {k}{n}$,
\begin{equation}
\label{eq:scaling}
{\cal K}_n(k) = \frac{k}{2 \pi n} \ K_{RMT}(\frac{2\pi n}{k}) \ .
\end{equation}
 Since $k$ is the parameter conjugate to the action differences, it follows
that the correlation length of the action spectrum is proportional to $\frac{1}{n}$. Moreover,
(\ref {eq:scaling}) shows that the classical correlations of actions corresponding to different periods are
identical up to a scaling, and universal, since they are expressed in terms of  a system independent function.
These predictions pertain exclusively to classical properties and, once they are derived within classical
mechanics, the road would be cleared  for a semiclassical derivation of the BGS conjecture.

The validity of {\it (i)} in the limit $\tau \rightarrow 0$ was first shown by M. Berry
\cite{berry}. In the present context, it follows from the fact that the actions are discrete
variables, and therefore, for sufficiently large $N$ the only correlations are the diagonal
ones.  Hence, for $k \rightarrow \infty$,
\begin{equation}
\label{eq:diago}
\langle K_{cl}(k;n)\rangle\ \rightarrow\ \sum_{a}  |g_a  A^{(n)}_a|^2\ \approx \ n
\langle g\rangle \
\end{equation}
where $g_a$ is the number of different orbits which share the same action (mostly
due to symmetry) and the summation is now restricted to trajectories with
different actions. The equality on the right hand side is derived by using the ergodic
sum-rule  \cite {hannayozorio} and denoting by $\langle g \rangle $ the average value
of the $g_a$.  Since $\langle g \rangle $ takes the value 1 for systems without time reversal
symmetry, and 2 for systems which are symmetric under time reversal, the leading terms in the RMT
results for the CUE and COE are indeed reproduced.

 To derive the quantum formfactor in a consistent semiclassical way, one has to consider the
classical formfactor for a fixed $n$, and change $k$ (or $N$) so that an interval of $\tau$ values
is scanned. However, to have meaningful pair correlations,  $N$ should be larger than 2, and therefore the
range of allowed $\tau$ values, is at most   $0< \tau <n/2$.  A further restriction on the range of $\tau$ is
due to the limited semiclassical accuracy, which  is of the order of a mean spacing, \cite
{accuracy}. In other words, the semiclassical trace formula may have its poles away  from  the unit circle,
thus replacing the sharp $\delta$ functions in (\ref {eq:qmdens}), by spikes with a finite width, of the order
of the mean spacing.  Therefore, the regime $\tau > 1$ is not expected to be accessible to the semiclassical
approximation, and thus, the range of applicability reduces further to the domain $n<N$. It is important to
note at this point that the required order of operations (keeping $n$ fixed and taking $N \rightarrow \infty$)
is consistent with the correspondence principle and the rules of quantum mechanics where the limit $\hbar
\rightarrow 0$ is taken at a fixed value of the classical parameters.

 In the present work we shall study the Baker map, which will be introduced in the next section. It is a simple
dynamical system, for which various properties can be derived analytically, yet it carries  the full
complexity of the action spectrum, and therefore it is appropriate as a paradigm.  In section (\ref {sec:density})
the spectrum of the actions will be discussed from various points of view, and on the different relevant scales. The
expression of the actions in terms of the symbolic codes of the periodic orbits will play a central role.  Mapping
the computation of the spectral density onto an Ising model enables us to introduce a few approximations
which enable the derivation of some properties of the action density (\ref {sec:Ising}). The classical pair
correlations  are discussed in section (\ref{correlations}). We study first the
$m$'th neighbor distributions. We show numerically that as long as $m$ is smaller than the action correlation length
$1/n$ measured in units of the mean action spacing, the $m$'th neighbor distributions are essentially Poissonian.
Turning to the formfactor, we discuss alternative smoothing methods  and relate them to the
symbolic codes. We finally show that the classical correlations reproduce the expected RMT behavior as was
conjectured in the general discussion presented above. We conclude the paper by a summary, where the connection
between the present  and previous works is discussed.

\section{The Baker map}

\subsection{The mapping}
\begin{figure}[h]
\centerline{\psfig{figure=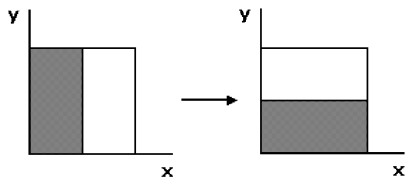,width=6cm}} \vspace{-7mm}
\caption{The baker map} \label{fig:baker}
\end{figure}

The \emph{Baker} map is one of the most simple examples of chaotic maps~\cite{saraceno},~\cite{balazsvoros},~\cite{saracenovoros}.
It is an area preserving map of the unit square onto itself defined by
\begin{equation}
\label{map}
x'=2x-[2x]\hspace{3mm} ;\hspace{3mm} y'=\frac{1}{2}y+\frac{1}{2}[2x]
\end{equation}
The stretching in the $x$ direction and the squeezing in the $y$ direction are
responsible for the hyperbolic character of the map, while the ``cutting and putting on
top'' gives the mixing property. Figure \ref{fig:baker}. shows one iteration of the map.

The action of the map is easily translated to a \emph{Bernoulli} shift:
Every phase space point is presented in a binary basis
\begin{eqnarray*}
x&=&\sum_{i=1}^{\infty}a_i2^{-i}=0\cdot a_1a_2a_3 ...\\
y&=&\sum_{i=1}^{\infty}b_i2^{-i}=0\cdot b_1b_2b_3 ...
\end{eqnarray*}
with $a_i,b_i\in\{0,1\}$.
The dynamics is given by shifting the binary point to the right when the two fractions
 are put back to back
\begin{eqnarray*}
...b_3b_2b_1\cdot a_1a_2a_3 ... &\longrightarrow& ...b_3b_2b_1a_1\cdot a_2a_3 ...\\
x=0\cdot a_1a_2a_3 ...          &\longrightarrow&     x'=0\cdot a_2a_3 ...\\
y=0\cdot b_1b_2b_3 ...          &\longrightarrow&     y'=0\cdot a_1b_1b_2b_3 ...
\end{eqnarray*}

\subsection{Periodic orbits and codes}
\label{subsect:po-codes}
A $n$-periodic point is represented as an infinite repetition of a finite binary
string with $n$ entries $\nu=(\nu_1,\nu_2,....\nu_n) , \nu_i\in\{0,1\}$.
The cyclic permutations of $(\nu_1....\nu_n)$ represent the periodic points which constitute
the periodic orbit.   The number of  $n$-periodic orbits  is  $\approx 2^n/n$.
The phase space coordinates of a $n$-periodic point can be written in term of the code
\begin{eqnarray}
x_i=\frac{1}{1-2^{-n}}\sum_{j=1}^n\nu_{i+j-1}2^{-j} \ \ ; \  \
y_i=\frac{1}{1-2^{-n}}\sum_{j=1}^n\nu_{i+j-1}2^{-n+j-1}
\end{eqnarray}
As $n\to\infty$ periodic points fill phase space densely and uniformly.

The symbolic dynamics introduced above is based on the partition of the unit square into two equal rectangles
along the line $x=\frac{1}{2}$. The sequence of binary symbols indicates the order by which the orbit visits
the rectangles.  Alternative codes can be generated by partitioning the unit square along the lines
$x_j= j \frac{1}{2^r}$, with $r>1$ integer, and  $1\le j \le 2^r-1$. A symbolic code consisting
of the $2^r$ symbols $0,1,\cdots,2^r-1$ indicates the order by which the  periodic orbit visits the rectangles.
The translation of a binary sequence to the $2^r$ code is done by
considering successive  sequences of $r$ binary symbols as integers in $[0,1,\cdots,2^r-1]$. (Example, the
binary code $\{ 001110101 \}$ is translated to the $2^{r=3}$ code $\{ 137652524 \}$).
Increasing $r$  the code becomes more informative, because it locates the rectangles with an accuracy $2^{-r}$
along the $x$ axis. However, this is achieved at a cost:  not all sequences of symbols are allowed. They are
restricted by a Markovian grammar with a $2^r \times 2^r$ connectivity  matrix with only two non vanishing entries
per row.
\begin{equation}
\label{C_r}
C^{(r)}_{j\ j'}=  \left\{\begin{array}{ll}\ 1  & \mbox{for $(j'-2j){\rm mod}\ 2^r \in \{ 0,1 \} $ }\vspace{2mm}\\
\ 0 & \mbox{otherwise}
\end {array}
\right.
\end{equation}
The refined codes will be used in the sequel.

\subsection{Symmetries}
\begin{figure}[h]
\centerline{\psfig{figure=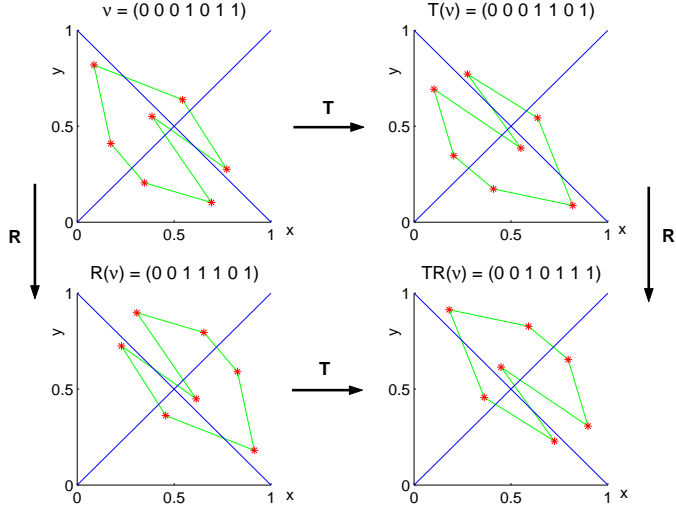,width=9cm}}
\vspace{-3mm}
\caption{Symmetry related periodic orbits}
\label{fig:bakersym}
\end{figure}

The mapping possess two discrete symmetries ~\cite{saraceno}. The first is a space reflection symmetry
\[ \hat{R} \ : \  x\longrightarrow1-x\hspace{1cm}\mbox{ and }\hspace{1cm}y\longrightarrow1-y\]
geometrically it is a double reflection about both the $y=1-x$ diagonal and the $y=x$ diagonal, it manifests itself on the
 code by $\hat{R}(\nu_i)=1-\nu_i$.
The second is  time reversal
\[ \hat{T} \ : \  x\longrightarrow y\hspace{5mm}\mbox{;}\hspace{5mm}y\longrightarrow x\hspace{5mm}\mbox{ and
}\hspace{5mm}t\longrightarrow -t\]  where reversing the time means reversing the mapping. Geometrically it is a
reflection about the $y=x$ diagonal and  its action on the code of a periodic orbit is $\hat{T}(\nu_i)=\nu_{n-i+1}$.
Figure \ref{fig:bakersym}. shows the action of the symmetry transformations: $\hat{R}$,
$\hat{T}$ and $\hat{R}\,\hat{T}=\hat{T}\hat{R}$ on the code $\nu\!=\!(0\,0\,0\,1\,0\,1\,1)$.

\subsection{Generating function, action}
The action $s$ associated with a phase space point $(x,y)$ is defined as the generating function
 $F_1(x,x')$ of the mapping, with $x'=x'(x,y)$. Because of the fact that in the Baker map, the $x$ dynamics is
independent of  $y$, one has to derive the action by first extracting the generating function
$F_2(x,y')$ from the conditions:
\[\frac{\partial F_2(x,y')}{\partial x}=y \hspace{2cm}\frac{\partial F_2(x,y')}{\partial y'}=x'\]
and get $F_2(x,y')=2xy'-x[2x]-y'[2x]+const$. To obtain $F_1(x,x')$ a \emph{Legendre} transform is
 needed (see~\cite{ozoriosareceno})
\[F_1(x,x')=F_2(x,y')-x'y'=-x[2x] \ .\] \
One should emphasize that $F_1(x,x')$ is not a generating function of the mapping.
However choosing $F_1(x,x')$ as the action is consistent with the action obtained from the
semiclassical approximation of the quantum Baker map ~\cite{saracenovoros,ozoriosareceno}.
Adding to the action an integer valued function $I(x)$ does not affect any semiclassical calculation
(see~\cite{dittesdoron}). The choice $I(x)=[2x]$ renders the action invariant under the symmetries of
the mapping. Inserting an overall minus (a matter of convention) gives: $s(x,x^{\prime })\equiv
s(x)=(x-1)[2x]$.  We use the symbol $s$ to denote the actions because the phase-space area is $||{\cal A}|| =1$, so
the actions are properly normalized. Applying the above to the $i$'th
periodic point (or
\emph{segment}) of a periodic orbit $\nu $ of length $n$, yields $s(x_{i},x_{i+1})\equiv
s(x_{i})=(x_{i}-1)[2x_{i}]=(x_{i}-1)\nu _{i}$. The total action of the periodic orbit $\nu $ is the
sum of its {\em segment }actions

\begin{equation}
\label{sn}
s_n(\nu)=\sum_{i=1}^ns(x_i)=\sum_{i=1}^n(x_i-1)\nu_i \ .
\end{equation}
The action thus defined is invariant under space reflection $\hat{R}$, time reversal $\hat{T}$ and
 the baker transformation $\hat{\mathcal{B}}$ which is only a cyclic permutation of $(\nu_1....\nu_n)$ i.e.
\begin{equation}
\label{eq:symm}
s_n(\nu)=s_n(\hat{R}(\nu))=s_n(\hat{T}(\nu))=s_n(\hat{\mathcal{B}}(\nu)) \ .
\end{equation}
Since the periodic orbits $\nu$, $\hat{R}(\nu)$, $\hat{T}(\nu)$, $\hat{R}\hat{T}(\nu)$ are
 not identical in general, one expects a maximal symmetry degeneracy of $4$.

The action can also be expressed in a matrix form~\cite{dittesdoron2} which will be useful later.
Writing $\left<\nu\right|=Vec(\nu_1....\nu_n)$ and $\left|\nu\right>=Vec(\nu_1....\nu_n)^T$ then the
action is
\begin{equation}
\label{smat}
s_n(\nu)=\left<\nu\right|\hat{O}\left|\nu\right>-\left<\nu|\nu\right> \ ,
\end{equation}
where $\hat{O}$ is a $n\times n$ cyclic matrix,
with matrix elements:
\begin{equation}
\label{smat2}
O_{ij}=\frac{1}{2^n-1}2^{(i-j+n-1)\,{\mathrm{mod}}\,n} \ .
\end{equation}
This expression of the actions clearly shows that their values are restricted to integer multiples of $2^{-n}$.
This is the minimum separation in the action spectrum.

The Baker map is uniformly hyperbolic. The stability eigenvalues of all the $n$-periodic orbits are the same, namely
$\lambda_{\pm}=2^{\pm n}$, the corresponding eigenvectors are parallel to the $(x,y)$ axes, and the Maslov indices are
all null. These features bring  a large degree of simplification which is to be incorporated in the definition  of the
action density for the Baker map.

\section {The action density}
\label{sec:density}

\begin{figure}
\centerline{\psfig{figure=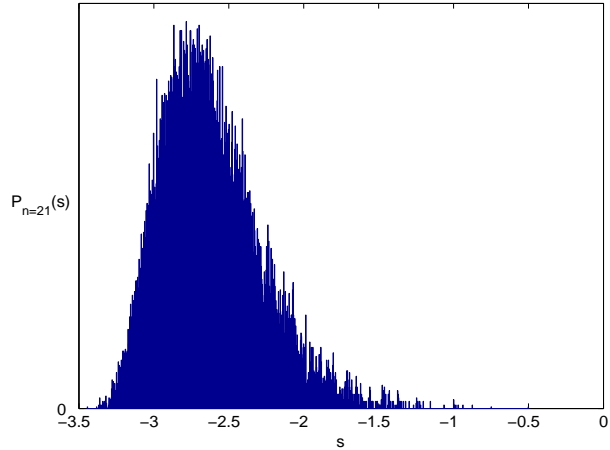,width=8cm,height=6cm}}
\vspace{-3mm}
\caption{Spectrum of actions, n=21}
\label{fig:sn21}
\end{figure}

The action density (\ref {eq:defdcl}) for the Baker map takes the form
\begin{equation}
\label{eq:bakerd}
d_{cl}(s;n) \ = \frac{1}{2^{\frac{n}{2}}-2^{-\frac{n}{2}}}\sum_ { \nu}  \delta (s- s (\nu)) \ ,
\end{equation}
where the summation extends over all the different vectors $\nu$. Note that in this way of writing, repetitions
are  properly weighted.  Since the weights in (\ref {eq:bakerd}) are all  positive, it is
convenient to normalize the density to unit integral. We shall denote the normalized density by $P_n(s)$, with
$\int P_{n}(s)ds=1$ (the subscript $cl$ is dropped since we shall deal exclusively with the
classical spectrum).
\begin{equation}
\label{Pns}
P_n(s)=\frac{1}{N_n}\sum_{a\in{\mathcal P}_n}\frac{1}{r_a} \delta(s-s_a )
\end{equation}
where ${\mathcal P}_n$  stands for the set of distinct $n$-periodic orbits and $r_a$ is the repetition number.
The normalization factor $N_n$, the number of $n$-periodic orbits,   tends to  $\frac{2^n}{n}$ in  the large $n$ limit.
An alternative expression is obtained by lumping together all the orbits which have the same action and the set
$\tilde{\mathcal P}_n$ consists of single representatives from each degeneracy set, and $g_a$ is the corresponding
degeneracy,
\begin{equation}
\label{Pnsprime}
P_n(s)=\frac{1}{N_n}\sum_{a\in \tilde{\mathcal P}_n}\frac{g_a}{r_a} \delta(s-s_a ) \ .
\end{equation}
 As defined in (\ref {Pns},\ref {Pnsprime}), $P_n(s)$ is a distribution. For some purposes, it is advantageous
to use a smooth version obtained  by convoluting (\ref {Pns},\ref {Pnsprime}) with a narrow window function. The
resulting smooth function will also be denoted as $P_n(s)$, and it is shown in figure  \ref {fig:sn21}. for $n=21$.

\begin{figure}[h]
\centerline{\psfig{figure=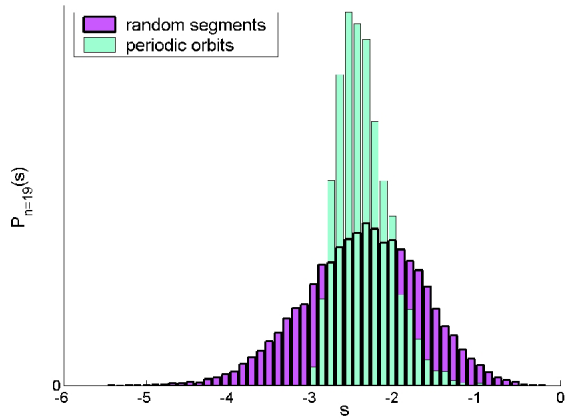,width=8cm,height=6cm}}
\vspace{-3mm} \caption{The density of randomly generated actions
for $n=19$, compared with the true density.} \label{fig:rand_seg}
\end{figure}

The simplest (but wrong) estimate for the action density is obtained by assuming that the segment actions $s(x_i) =
(x_i-1)\nu_i$ (\ref {sn}) are independent random variables which are uniformly distributed on the interval
$[-\frac{1}{2},0]$. This leads to a Gaussian distribution in the limit of large $n$, which has the same mean as the
 action distribution but otherwise  is quite different from it. Figure \ref{fig:rand_seg}. compares the
distribution computed numerically for a random choice of segments  ($n=19$),  with the
actual  action distribution (periodic orbits).

  We shall start the discussion of the action density by reviewing some general properties. In the next
subsection, we shall investigate the substructures in the spectrum, and show that it partitions naturally to families
which can be characterized in terms of symbolic codes.

\subsection {General properties}
\label{subsec:general}

The action distributions $P_n(s)$ is characterized by various scales whose dependence on $n$ will be
summarized below.

\begin{figure}[h]
\centerline{\psfig{figure=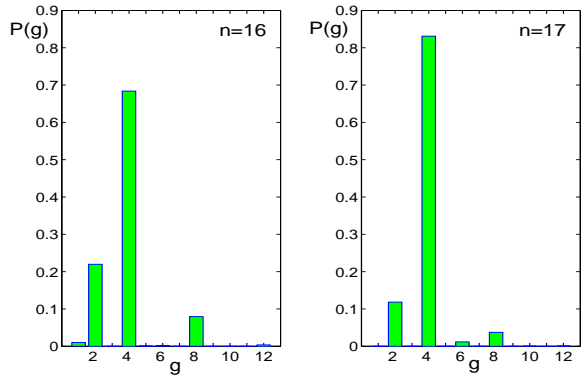,width=8cm,height=5cm}}
\vspace{-3mm}
\caption{The distribution of degeneracies for $n=16,17$.}
\label{fig:deg}
\end{figure}

\noindent \underline {Degeneracy} in the action spectrum is mainly due to the symmetries (\ref {eq:symm}).  As $n$
increases, a larger fraction  of the orbits are not self symmetric, and therefore the mean degeneracy is
$\langle g \rangle =4$.  Figure (\ref {fig:deg}) shows the degeneracy distributions for n=16,17. One can see that
 for these values of $n$ the degeneracy 2 and 1 still appear with appreciable frequency, and higher degeneracies whose
origin is number theoretical are also possible.

\noindent \underline {The lowest scale} is $2^{-n}$ since the actions are integer multiples of this interval.

\noindent \underline {The largest scale} is provided by the interval $I_n$ which supports the action distribution.
It follows directly from (\ref{smat}) that the maximum value of $s_n$ is $s_n^{max}=0$ while the
minimum is given by the action associated with the periodic orbit  $\nu=(0\  1\  0\  1\  0\  1\ \ldots)$.
\begin{equation}
\label{Nn}
s_n^{min}=- \left\{\begin{array}{ll}\frac{n}{6}  & \mbox{for $n$ even}\vspace{2mm}\\
\frac{n}{6}\left(1-\frac{1}{3n}\frac{2^n+1}{2^n-1}\right) & \mbox{for $n$
odd}\end{array}\right.
\end{equation}
Hence, to leading order $I_n= \frac{n}{6}$. This interval accommodates $N_n\approx \frac{2^n}{n}$ periodic
orbits, which are $\approx$ 4 times degenerate. It follows that

\noindent \underline{The mean spacing} is
    \begin{equation}
\label{meanspacing2}
\left<\Delta s^{(n)}\right> \, \simeq \, \frac{2}{3}\frac{n^2}{2^n}.
\end{equation}
This estimate reproduces very well the numerical simulations. The above result indicates that not all the integer multiples of
$2^{-n}$ on the action line are populated. Rather, the gaps are of order $n^2$. They cannot be seen in figure (\ref{fig:sn21}),
because the bin size used is too coarse.

To characterize the large scale features of the action distribution we quote explicitly its first 3 moments in the limit of large
$n$ (the proof of these relations is given in section (\ref {sec:Ising}) below) :
\begin{eqnarray}
\label{moments}
\left<s_{n}\right>\ &\longrightarrow& -\frac{n}{8}\nonumber  \\
{\mathrm{var}}(s_{n})\ &\longrightarrow&\ \ \frac{n}{192}\nonumber  \\
 \left<\left(s_{n}-\left<s_{n}\right>\right)^3\right>\ &\longrightarrow &\ \ \frac{n}{768}
\end{eqnarray}

\begin{figure}[h]
\centerline{\psfig{figure=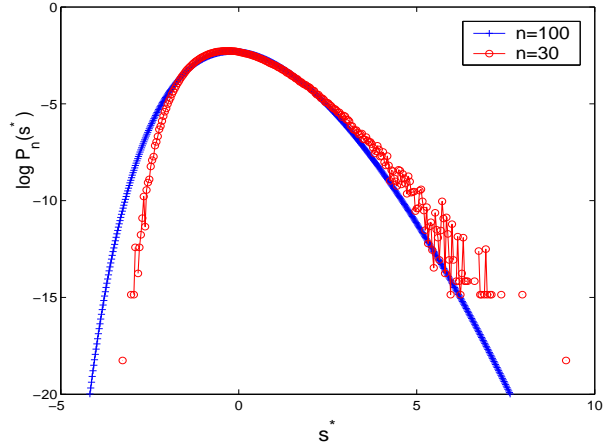,width=8cm,height=6cm}}
\vspace{-3mm} \caption{Scaled action densities for $n=30,100$.}
\label{fig:largenscal}
\end{figure}

To check the limiting action distribution, it is appropriate to examine its dependence on the scaled action:
\begin{equation}
\label{eq:sstar}
s^* =\frac {s- \left<s_{n}\right>}{  \sqrt{  {\mathrm{var}} (s_{n}) } }
\end{equation}
Sano \cite {sano} has recently shown that the scaled action density becomes Gaussian in the limit of large $n$. Our numerical
computations confirm that
\begin{eqnarray}
\label{momlarge}
&&\lim_{n \to \infty}\frac{\left<\left(s_n-\left<s_n\right>\right)^l\right>}{(l-1)!!n^{l/2}}=1\hspace{10mm
}\mbox{for $l$ even}\nonumber\\
&&\lim_{n \to \infty}\frac{\left<\left(s_n-\left<s_n\right>\right)^l\right>}{n^{l/2}}=0\hspace{10mm}\mbox{for $l$ odd} \ .
\end{eqnarray}
Figure (\ref {fig:largenscal}) compares the scaled distributions for $n=30,100$, (computed for the $r=5$
level of the ising model of section (\ref {sec:Ising}). One can clearly see that the distribution for the larger $n$ gets more
symmetric. The fact that the distribution of scaled actions tends to a  Gaussian does not imply that pair correlations do not exist.

\subsection {Families of actions and symbolic codes}
\label{subsect:fam-codes}

 Symbolic codes are naturally associated with a partition of phase-space. (See e.g., \cite {Sinailect}). The
actions, being functions of phase-space points are expressed in terms of the codes (\ref {smat}) and
therefore their classification into families which share certain properties is conveniently carried out in terms of
their codes.

\begin{figure}[h]
\centerline{\psfig{figure=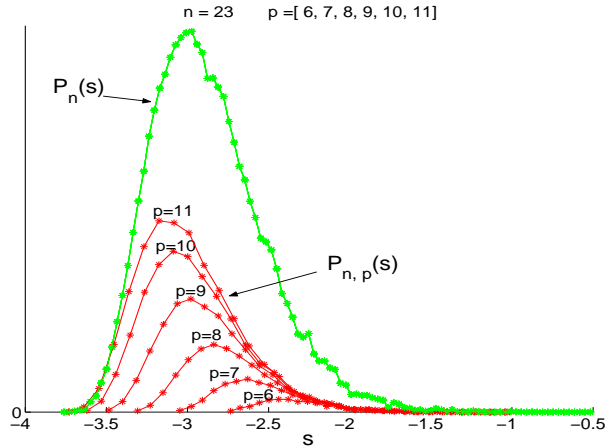,width=8cm,height=6cm}}
\vspace{-3mm} \caption{ distributions of actions according to
$p$-families.} \label{fig:sn23p}
\end{figure}

\begin{figure}[h]
\centerline{\psfig{figure=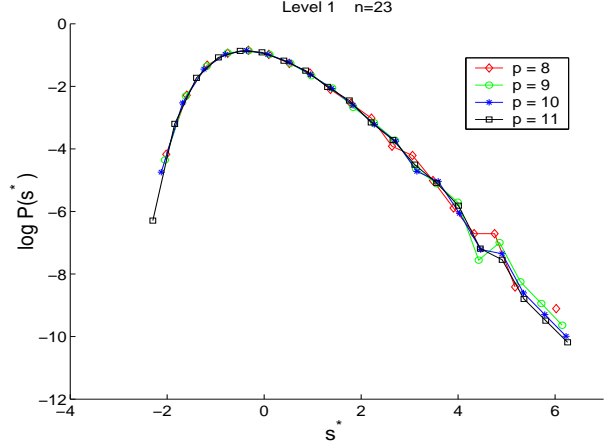,width=8cm,height=6cm}}
\vspace{-3mm} \caption{Scaling of the $p$-family distributions.}
\label{fig:snpscaled}
\end{figure}

 As was shown in section \ref {subsect:po-codes} the binary code $\{0,1\}$ is based on the partition of phase space
by a vertical line  at $x=\frac{1}{2}$ into the two rectangles which are associated to the codes
 $0$, and $1$.  Only the points of the trajectory which fall in the second rectangle
(code $1$) contribute to the action, and the increment to the total action per point is $x_i-1$. Denote by $p$ the
number of times the periodic orbit visits the second interval, or equivalently, the number of $\nu_i=1$ in the code.
The set of all the trajectories which have the same $p$ will be referred to as a ``$p$-family".  The action of
trajectories within a $p$-family cluster in substructures which are illustrated in figure \ref{fig:sn23p}. in terms of
the densities $P_{n,p}(s)$, for $n=23$ and a few values of $p$. The  $P_{n,p}(s)$  are narrower than $P_{n}(s)$, and
one can show that they  are centered about $-\frac{1}{2}\frac{p(n-p)}{n-1}$.  Scaling to unit variance and shifting to
their mean
$s$ value (figure \ref  {fig:snpscaled}.), the densities $P_{n,p}(s)$ collapse to a single function which is very
similar to the scaled total density. Thus, the families enable a study of the spectrum of actions with a
finer resolution, with an approximate scaling similarity of their densities.

The partition of the actions into $p$-families is just the first level in a systematic procedure which enables the
sorting of the actions according to their codes. The higher the level, the more refined are the resulting families.
The higher level families still maintain the scaling similarities which was discussed for the $p$-families.

The $r$'th level families, (with $r<n$)  are obtained by  a partition of phase space into $2^r$ identical
rectangles of unit hight, and with $\frac {j-1}{2^r}\le x < \frac {j}{2^r}$, for all $1\le j \le 2^r$.
The family consists of all the $n$-periodic orbits which go through each of the $2^{r-1}$ rectangles in the rightmost
half of the unit square the same number of times. Given a
$n$-periodic orbit, we denote by  $p_l^{(r)}$, the number of times it goes through the
$l$'th rectangle with $\frac{1}{2}+\frac{l-1}{2^r}\le x < \frac{1}{2}+\frac{l}{2^r}$,  and  $1\le l \le 2^{r-1}$.
A family of the $r$'th level is characterized by
the set of $2^{r-1}$ numbers
\begin{equation}
\label{eq:Pcode}
{\mathbf{P}}^{(r)}=\left(p_1^{(r)},p_2^{(r)},\ldots,p_{2^{r-1}}^{(r)}\right)\ .
\end{equation}

One can get the ${\mathbf{P}}$ code by direct inspection of the binary code $\nu$.  $p_j^{(r)}$ is the number of
times the  $r$ digits string, beginning by 1 and terminating by $j\!-\!1$ written in a binary basis, occurs in $\nu$
i.e
\begin{equation}
 p_j^{(r)}(\nu)\,\equiv\, \#\,(\underbrace{1 \, \overbrace{0\, 0\, \cdots 0 1\cdots\cdots\cdots}^{j\!-\!1\,
\mathrm{in\,
 binary}}}_{r\,\mathrm{digits}}\,)\hspace{5mm}\mathrm{in}\,\ \nu  \ .
\end{equation}
One can easily check that the $p$ families defined at the beginning of this sections correspond to $r=1$, with
$p_1^{(1)}\equiv p$. The $r$
 and the  $r+1$ partitions are related by
\[  p_j^{(r)}=p_{2j}^{(r+1)}+p_{2j-1}^{(r+1)}  \]
 Also,
\[\forall r : \hspace{5mm}\sum_{j=1}^{2^{r-1}}p_j^{(r)}=p_1^{(1)}  \equiv p\ .   \]
 The mean action for a  ${\mathbf{P}}^{(r)}$ family can be written as
\begin{equation}
\label{snqr}
{\hspace {-6mm}} s_n^{(r)}=\sum_{j=1}^{2^{r-1}}p_j^{(r)}\left(-\frac{1}{2}+\frac{2j-1}{2^{r+1}}\right) \ =\
-\frac{p}{2}(1+\frac{1}{2^r})\ +\
\frac{p}{2^r} \ \frac{\sum_{j=1}^{2^{r-1}}j\  p_j^{(r)}}{\sum_{j=1}^{2^{r-1}} p_j^{(r)}} \ .
\end{equation}
Actions which belong to the same family cluster about this mean value within an interval of order  $\frac{\sqrt
n}{2^r}$.

The partition into families  break the $\hat R$ symmetry in the sense that symmetry conjugate pairs of periodic
orbits occur in different families. This is clear from the observation that the number of $1$ in the two conjugate
codes are $p$ and $n-p$.    For
$r\ge 3$,  periodic orbits which are related by
$\hat T$   can be assigned to  different families.  (Example: the periodic orbits which are represented by the binary
codes  $(001011)$ and $(001101)$ are  $\hat T$ conjugate but they appear in the $r=4$ families $(01011000)$ and
$(01100100)$, respectively).  To avoid this problem, one should partition the actions which correspond to the set
$\tilde{\mathcal P}_n$, where only one representative of each degeneracy class is included (\ref {Pnsprime}).

 An alternative partitioning of the actions in families can be defined in the following way. Given a binary code
$\nu$, let $q_1$ be the number of $1$ in $\nu$, let $q_2$ be the number of sequences $\{11\}$ in $\nu$, and in
general, $q_r$  is the number of sequences of $r$ consecutive $1$ in $\nu$. (Clearly $q_1=p$). The $r'th$ level family
consists of all the
$n$-periodic orbits which have the same ${\mathbf{Q}}^{(r)}$ code,
\begin{equation}
\label{eq:Qcode}
{\mathbf{Q}}^{(r)}=\left(q_1^{(r)},q_2^{(r)},\ldots,q_{r}^{(r)}\right)\ .
\end{equation}
This partitioning has the advantage that $\hat T$-conjugate orbits are always in the same family,
(however, $\hat R$-conjugate orbits are in different families). We found this partitioning more convenient for
numerical simulations. The two methods of partitioning, at the same $r$ level, have in common the same resolution
$2^{-r}$.

    The partition of the actions of $n$-periodic orbits into families provides a very useful tool for the analysis of
the spectral correlations, especially because the parameter $r$ which is at our disposal, determines the resolution at
which we wish to interrogate the spectrum.  Given $r$, most pairs of actions at a distance less than $2^{-r}$ belong
to the same family, and hence,  {\it intra} family investigations are sufficient for the study of smaller  action
differences. However, for larger scales,
 all actions in a family can be lumped together, and the large range correlations are expressed through
the study of {\it inter} family correlations. We shall take both approaches when we discuss the two-point
correlations in the action spectra.

\subsection{An Ising model approach}
\label{sec:Ising}

In the present section we shall introduce a method to investigate the action spectra on scales which are larger
than a given level of resolution. For this purpose, it is convenient to study the
 Fourier transform of the action distribution, $\hat{P}_n(k)$,
\begin{equation}
\label{pnkmodel}
 \hat{P}_n(k) =\frac{1}{2^n}\sum_{\{\nu\}}e^{iks_n(\nu)}
 =\frac{1}{2^n}\sum_{\{\nu\}}e^{ik\left(\sum_{i,j}^n\nu_iO_{ij}\nu_j-\sum_j^n\nu_j^2\right)}
\end{equation}
where the sum is over all possible configurations of the code $\nu=(\nu_1,\nu_2,....\nu_n)$, $\nu_i\in\{0,1\}$.
 The action is given explicitly by

\begin{equation}
s(\nu_n)=
\frac{1}{1-2^{-n}}\sum_i\left[\frac{1}{2}\nu_i^2+\frac{1}{4}\nu_i\nu_{i+1}+
\frac{1}{8}\nu_i\nu_{i+2}+\cdots\right]
-\sum_i\nu_i^2 \ .
\end{equation}
Formally~(\ref{pnkmodel}) is the partition sum of a one dimensional \emph{Ising} model on a circular lattice with
exponentially decreasing interactions and an imaginary temperature. It can be evaluated using the standard
transfer matrix method. Using this method we can approximate $\hat{P}_n(k)$ by truncating the interaction at any
desired range $r$, $(1\le r \le n-1)$. Thus, $r=1$ is the nearest neighbors approximation, $r=2$ is the next to
nearest neighbors approximation etc...\hspace{3mm}. When $r=n-1$ we regain the full range of interactions.
The original effective hamiltonian is invariant under space reflection $\hat{R}$ and time reversal $\hat{T}$. The
invariance under $\hat{T}$ is due to the cyclic property of $\hat{O}$ and hence, is maintained for any truncated
version. This is not the case however for the space reflection $\hat{R}$ symmetry. In order to preserve $\hat{R}$
invariance one has to rescale the strength of the term $\sum_j\nu_j^2$ which was introduced for $\hat{R}$ invariance
of the full action~(\ref{sn}). For a given range $r$ of the approximation, the action takes the form

\begin{equation}
\label{snr}
s_{n,(r)}(\nu)=\frac{1}{1-2^{-n}}\,\sum_{\xi=0}^r\,\sum_{i=1}^n\,\frac{1}{2^{(\xi+1)}}\nu_i\nu_{i+\xi}
-\gamma_n(r)\,\sum_{i=1}^n\nu_i^2
\end{equation}

The symmetry restoring coefficient $\gamma_n(r)$ is calculated by demanding that for the $r^{th}$
order approximation the action be $\hat{R}$ symmetric i.e. $s_{n,(r)}(\nu)=s_{n,(r)}(\hat{R}(\nu))$ and is given
by

\begin{equation}
\gamma_n(r)=\frac{1}{1-2^{-n}}\sum_{\xi=0}^r\,\frac{1}{2^{(\xi+1)}}=\frac{\,1-(\frac{1}{2})^{r+1}}
{1-(\frac{1}{2})^n}
\end{equation}
One recovers $\gamma_n(r)=1$ for the full range interactions $r=n-1$.

This completes the definition of the $r$'th level approximants, and because of the exponential decreasing strength of the
interactions we can expect that any quantity  computed at the $r$'th level, will converge exponentially fast to its
full range interaction value.

Note: for the rest of this section we omit the  factor
$\frac{1}{1\!-\!2^{-n}} \approx 1$ which multiplies $s_n$.

We express $\hat{P}_n(k)$, for  a given range $r$ as:
\begin{equation}
\hat{P}_n^{(r)}(k)=\frac{1}{2^n}{\mathrm{tr}}\left \{[T^{(r)}(k)]^n \right \}  \ ,
\end{equation}
where the transfer matrix $T^{(r)}(k)$ is given by

\begin{eqnarray}
\label{Tmatrix}
{\hspace {-10mm}}
 T^{(r)}(k)=  \left(\!
\begin{array}{cccccccc}
1&1&0&&&&\cdots&0\vspace{+1mm}\\
0&0&1&1&0&&\cdots&0\vspace{+1mm}\\
.&&&&&&&.\vspace{-3mm}\\
.&&&&&&&.\vspace{-3mm}\\
.&&&&&&&.\vspace{-3mm}\\
.&&&&&&&.\vspace{+1mm}\\
0&0&&&\cdots&0&1&1\vspace{+1mm}\\
\!\beta^{2^r\!\!-\!1}\!&\!\beta^{2^r\!\!-\!2}\!&\!0&&&&\cdots&0\vspace{+1mm}\\
0&0&\!\beta^{2^r\!\!-\!3}\!&\!\beta^{2^r\!\!-\!4\!}&\!0&&\cdots&0\vspace{+1mm}\\
.&&&&&&&.\vspace{-3mm}\\
.&&&&&&&.\vspace{-3mm}\\
.&&&&&&&.\vspace{-3mm}\\
.&&&&&&&.\vspace{+1mm}\\
0&0&&&\cdots&0&\beta&1\vspace{+1mm}
\end{array}
\right)\ \ ; \
\beta=e^{-i\frac{k}{2^{r+1}}}
\end{eqnarray}
Since  $T$ depends on powers of $\beta$, ${\mathrm{tr}}\left \{[T^{(r)}]^n \right \}$ is  a polynomial in $\beta$ with real
coefficients
 $A_j$
\begin{equation}
\frac{1}{2^n}{\mathrm{tr}}\left \{[T^{(r)}(k)]^n \right
\}=\sum_{j=0}^{N(n,r)} A_j\beta^j=\sum_{j=0}^{N(n,r)}A_je^{-ik\frac{j}{2^{r+1}}}
\end{equation}
where $N(n,r)$ is the degree of the polynomial. Transforming back to $P_n(s)$ yields
\begin{equation}
\label{pnrs}
P_n^{(r)}(s)=\sum_{j=0}^{N(n,r)}A_j^{(r)}\,\delta(s+\frac{j}{2^{r+1}})
\end{equation}
and
\begin{equation}
\sum_{j=0}^{N(n,r)}A_j^{(r)}=1 \ .
\end{equation}
The $r$'th approximant to the action spectrum consists of $N(n,r)$ equally spaced actions $s_j\!=\!-\frac{j}{2^{r+1}}$,
 each weighted by a normalized weight (or probability) $A_j^{(r)}$.   The
$A_j$'s contain, together with the spacing $\delta s(n,r)\equiv\frac{1}{2^{r+1}}$ , all the information about
the statistical properties of the spectrum at the resolution $\frac{1}{2^{r+1}}$ . Increasing the range $r$ to $r+1$
results in approximately doubling the number of actions, in addition to distributing them on a lattice with half the
spacing. In the limit $r\rightarrow n-1$ (full range interactions) the lattice spacing becomes $\frac{1}{2^n}$ (or
$\frac{1}{2^n-1}$ taking into account the factor $\frac{1}{1-2^{-n}}$ that was neglected). This observation enables us to
connect the present approach with the partitioning of the actions to families according to their codes. The truncation of
the interaction at the $r$'th level is approximately equivalent to replacing the actions of all the members of a family by
their average value given by (\ref {snqr}). Thus, the coefficients $2^n A_j^{(r)}$ approximate the sum of the
cardinalities of  all the families whose average action is $s_j\!=\!-\frac{j}{2^{r+1}}$.

The  moments of the distribution are computed using
\begin{eqnarray}
\label{moments}
\hspace{-5mm}\left< s_{n,(r)}^m \right>\!&=&\!(-i)^m \left. \frac{\partial^m P_n^{(r)}(k)}{\partial k^m}
 \right|_{k=0}\nonumber\\
&=&\!\left(\frac{-1}{2^{r+1}}\right)^m\,\sum_{j=0}^{N(n,r)}A_j^{(r)}j^m \ .
\end{eqnarray}


The two lowest level approximants can be solved analytically.

\noindent \underline{$r=1$} : The transfer matrix  (\ref{Tmatrix}) for this ``nearest neighbors interaction" approximation
is
\begin{equation}
T^{(r=1)}(k)=
\left(
\begin{array}{cc}
1&1 \\
\beta&1
\end{array}
\right)
\hspace{5mm};\hspace{5mm}\beta=e^{-i\frac{k}{4}}
\end{equation}
Its eigenvalues are $\lambda _{\pm }=1\pm
\sqrt{\beta }$ which yields
\begin{eqnarray}
\hat{P}_{n}^{(r=1)}(k)\!&\!=\!&\!\frac{1}{2^{n}}\left[\left(1+\sqrt{\beta }\right)^{n}+\left(1-\sqrt{\beta }\right)^{n}\right]\nonumber\\
\!&\!=\!&\!\frac{2}{2^{n}}\sum_{m=0;even}^{n}\!\!\left(\!\!\begin{array}{c}n\\m\end{array}\!\!\right)e^{-ik\frac{m}{8}} \
,
\end{eqnarray}
and,
\begin{equation}
\label{pnsr1}
P_{n}^{(r=1)}(s)=\frac{2}{2^{n}}\left(\!\!\begin{array}{c}n\\-8s\end{array}\!\!\right) \ ,
\end{equation}
with
\[ s \in\left\{\,0\,,\,-\frac{1}{4}\,\,,\,-\frac{2}{4}\,\,,\,-\frac{3}{4}\,\,,\,\ldots\,,\,-\frac{1}{4}\left[\frac{n}{2}\right]\,\right\}\]
This distribution limits to a  Gaussian distribution for large $n$.
The mean and the variance are given by
\[\left<s_{n,(r=1)}\right>=-\frac{n}{16}\ \ \ ; \ \ \
 {\mathrm{var}}(s_{n,(r=1)})=\frac{n}{256}\ \]

\noindent \underline {$r=2$}  : The transfer matrix in this ``next to nearest neighbors interaction" approximation is
\begin{equation}
T^{(r=2)}(k)=
\left(
\begin{array}{cccc}
1&1&0&0 \\
0&0&1&1\\
\beta^3&\beta^2&0&0\\
0&0&\beta&1
\end{array}
\right)
\hspace{1mm};\hspace{1mm}\beta=e^{-i\frac{k}{8}}
\end{equation}
The eigenvalues of $T^{(r=2)}$ are
\[\lambda _{1\pm }=\frac{1}{2}\left(1-\beta\pm\sqrt{\Delta_1 }\right) \ ; \
 \lambda _{2\pm }=\frac{1}{2}\left(1+\beta\pm\sqrt{\Delta_2 }\right)\]
with   $\Delta_1=1+2\beta-3\beta^2\hspace{1mm};\hspace{1mm}\Delta_2=1-2\beta+5\beta^2\hspace{5mm}$.
This gives:
\begin{eqnarray}
\label{pnkr2}
\hspace{-15mm}&&\hat{P}_{n}^{(r=2)}(k)=\\
\hspace{-15mm}&&\frac{2}{2^{2n}}\!\sum_{m=0;even}^{n}
\sum_{j=0}^{n-m}\sum_{u=0}^{\frac{m}{2}}\sum_{t=0}^{\frac{m}{2}-u}\left[\left(\!\!\begin{array}{c}n\\m\end{array}
\!\!\right)\left(\!\!\begin{array}{c}n-m\\j\end{array}\!\!\right)\right.
\left(\!\!\begin{array}{c}\frac{m}{2}\\u\end{array}\!\!\right)\left.\left(\!\!
\begin{array}{c}\frac{m}{2}-u\\t\end{array}\!\!\right)C_{jtu}e^{-ik\frac{j+t+2u}{8}}\right]\ , \nonumber
\end{eqnarray}
with
\[C_{jtu}=2^t\left\{3^u(-1)^{j+u}+5^u(-1)^t\right\}\ . \]
The action distribution is
\begin{eqnarray}
\hspace{-13mm}&&P_{n}^{(r=2)}(s)=\\
\hspace{-13mm}&&\frac{2}{2^{2n}}\!\sum_{m=0;even}^{n}\sum_{j=0}^{n-m}\sum_{u=0}^{\frac{m}{2}}\left[
\left(\!\!\begin{array}{c}n\\m\end{array}\!\!\right)\left(\!\!\begin{array}{c}n-m\\j\end{array}\!\!\right)
\right.
\left(\!\!\begin{array}{c}\frac{m}{2}\\u\end{array}\!\!\right)\left.
\left(\!\!\begin{array}{c}\frac{m}{2}-u\\-8s-2u-j\end{array}\!\!\right)C_{ju}(s)\right] \nonumber
\end{eqnarray}
with
\[C_{ju}(s)=2^{-8s-2u-j}\!\left\{3^u(\!-\!1\!)^{j+u}\!+\!5^u(\!-\!1\!)^{-8s-j}\right\} \ .\]
The action takes the values
\[ s \in\left\{\,0\,,\,-\frac{1}{8}\,\,,\,-\frac{2}{8}\,\,,\,-\frac{3}{8}\,\,,\,\ldots\,,\,-\frac{n}{8}\,\right\}\ .\]
and the summation coefficients $u, j, m$ must fulfill the conditions
\[2u+j\leq -8s \leq u+j+m/2\]
to keep the binomial coefficients well defined. The $r\!=\!2$ distribution is not Gaussian, its mean, variance and third
moment can be computed by a straight forward but cumbersome calculation.
\[\left<s_{n,(r=2)}\right>=-\frac{n}{16}(1+\frac{1}{2})
 \ \ \  ; \ \ \  {\mathrm{var}}(s_{n,(r=2)})=\frac{n}{256}(1+\frac{1}{4})\]
\[\left<\left(s_{n,(r=2)}-\left<s_{n,(r=2)}\right>\right)^3\right>=\frac{3n}{4096}\]

We were unable to compute analytically the action distributions at higher levels, since this involves finding the
eigenvalues of the $2^r$ dimensional transfer matrices.  However, the computation can be performed using computer
codes which perform algebraic manipulations.  For any  $r$ and $\beta$ one can compute
${\mathrm{tr}}\left \{[T^{(r)}(k)]^n \right \}$   by   raising $T^{(r)}$ to the power $n$ and taking the trace. Using
Newton's identities, one can compute the coefficients of the characteristic polynomial of  $T^{(r)}$ from the traces
of its lower powers, ${\mathrm{tr}}\left
\{[T^{(r)}(k)]^l \right \} \ ,\ l=1,\cdots,2^r$. The  traces of higher powers can then be expressed recursively in terms of these
coefficients. This way we were able to perform computations up to the level $r=9$, without reaching the limits of our computer
resources.

At the beginning of the chapter we argued that due to the exponential decrease of the
interactions,  computing any quantity at a given
level $r$, converges exponentially fast to  the
full $r=n-1$ calculation.  Thus, it can be expected that any quantity of
interest will be given by a geometric series in $2^{-r}$. Knowing the first two levels
($r=1,2$) may allow us to extrapolate, and obtain the leading terms for $r=n-1$.
Applying this strategy for the first three  moments, we obtained
the expression
\begin{equation}
\label{mom1eq}
\left<s_{n,(r)}\right>=-\frac{n}{8}\left(1-\frac{1}{2^r}\right)\longrightarrow -\frac{n}{8}
\end{equation}
\begin{equation}
\label{mom2eq}
{\mathrm{var}}(s_{n,(r)})=\frac{n}{192}\left(1-\frac{1}{4^r}\right)\longrightarrow \frac{n}{192}
\end{equation}
\begin{eqnarray}
\label{mom3eq}
\hspace{-6mm}\left<\left(s_{n,(r)}-\left<s_{n,(r)}\right>\right)^3\right>=
\frac{n}{768}\left(1-\frac{1+3r}{4^r}\right)\ \longrightarrow \frac{n}{768}
\end{eqnarray}
These expressions were checked by comparing them to the computer aided, large  $r$ calculations.

 We shall return to the results of the Ising model when we discuss the pair correlations in the action spectrum
which is the subject of the next section.


\section{Pair correlations in the actions spectrum}
\label{correlations}
 So far we discussed the {\it distribution} of actions. Now we turn to study their {\it  pair correlations}
which is the main issue of the present work. We examine whether the classical actions
spectrum exhibits correlations, and  whether these correlations conform with the semiclassical
theory which was presented in section (\ref{introduction}).

There are two distinct length scales in any discrete and finite spectrum: the width or
support of the distribution on the largest scale, and the nearest neighbors spacing on the
smallest scale. The fact that the support is finite induces a trivial
correlation: the probability to find two points separated by a distance larger than the
range of the support is zero. The nearest neighbors scale is of great interest, in particular for
``rigid" spectra such as the  quantum spectra of  classically chaotic systems. They exhibit \emph{level
 repulsion}  as predicted by $RMT$ i.e. the probability to find two nearest neighbors at distance
$\delta$ vanishes as $\delta \rightarrow 0$. We have already hinted in section (\ref{introduction}) that the
pair correlations in the action spectrum is expected on a scale which is much larger that the mean spacing.

We shall perform the analysis of pair correlations in several steps.
In the first we shall examine the correlations on the scale of the mean level spacing. We shall show that
the $m$'th nearest neighbor spacing distribution are consistent with the corresponding Poisson distributions,
as long as $m$ is smaller than the correlation length measured in units of the mean spacing. We shall then study the
formfactor of the pair correlation function, and compute it in various ways. The main result of this
investigation is that the correlations exist, and their scaling with $n$ agrees with the semiclassical expectations.

\begin{figure}[h]
\centerline{\psfig{figure=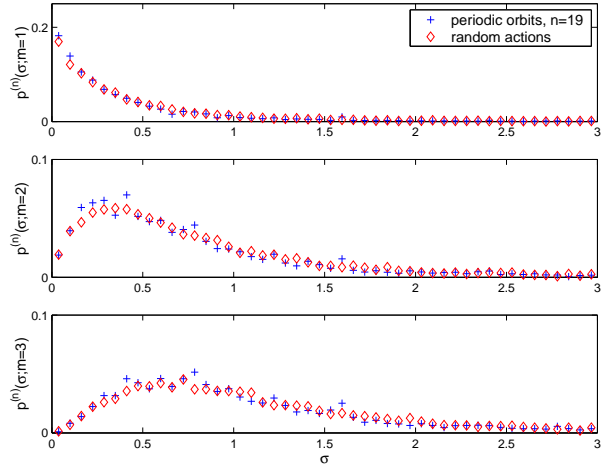,width=8cm}}
\vspace{-3mm}
\caption{The first, second and third neighbors distribution
($n=19$)} \label{fig:nndist}
\end{figure}

\subsection{The $m$'th-neighbors spacings distributions}
\label{nndistribution}

The $m$'th-neighbors spacing distribution $p^{(n)}(\sigma;m)$,  is a straight-forward
generalization of the widely used nearest neighbors spacing distribution. To define this distribution, the
action spectrum is ordered so that $s_i < s_{i+1}$, degeneracy sets are represented by a single value and the
spectrum is normalized by the  nearest neighbor spacings
\begin{equation}
\label{meanspacing2}
\left< \Delta s^{(n)}\right> \, \simeq \, \frac{2}{3}\frac{n^2}{2^n} \ .
\end{equation}
Then, $p^{(n)}(\sigma;m)$ is the probability that $\frac{s_{i+m}-s_{i}}{\left< \Delta s^{(n)}\right>}$
takes the value $\sigma$.

As a reference spectrum we generated  numerically a Poissonian spectrum. It is randomly chosen from a
Gaussian probability density which has the same mean, variance  and
degeneracy structure as the action spectrum for the $n$ value of interest.
In the upper frame of figure \ref{fig:nndist}.  the nearest neighbors distribution, $p^{(n)}(\sigma;m=1)$, for $n=19$ is
plotted together with the  random (Poissonian) nearest neighbors distribution. The
similarity between the two distribution, which persists over 6 orders of magnitude,  provides strong evidence in favor of the
claim that on the mean spacing scale, the action spectrum is statistically random.

This finding is further corroborated in the lower frames of figure ~\ref{fig:nndist}.  were  the spacing distributions for
$m=2,3$ are compared with the corresponding random distributions.

To go even further in $m$, we studied numerically the summed distributions
\begin{equation}
p^{(n)}_M(\sigma) = \sum_{m=1}^M p^{(n)}(\sigma;m) \ ,
\end{equation}
which approximates the two point correlation function $R_{2;n}(\sigma)=p^{(n)}_{\infty}(\sigma)$ in the
range $0 < \sigma < M$.

 \begin{figure}[h]
\centerline{\psfig{figure=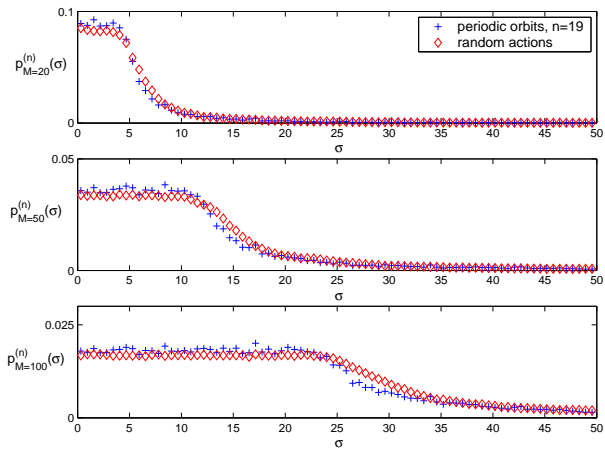,width=8cm}}
\vspace{-3mm}
\caption{Integrated distributions with $(M=20,50,100)$ ,
($n=19$)} \label{fig:pnnM_n19}
\end{figure}

Figure~\ref{fig:pnnM_n19}. shows $p_M^{(n)}(\sigma)$ for $M=20,50,100$, ($n=19$) together with the
corresponding random distributions.  These values $M$ should be compared with the correlation length
expected to be of order
\begin{equation}
\sigma_{corr} \approx \frac{1/n} { \left< \Delta s^{(n)}\right>} = 3 \frac{2^{n-1}}{n^3} \ \ \ (\ \approx
 \ 90 \  {\rm for} \ n\ = \ 19\ )
\ .
\end{equation}

This numerical investigation clearly demonstrates that the
systematic deviation between the random and the actual
distributions increases as $M$ approaches the correlation length.


\subsection{The classical spectral formfactor }
\label{subsect:ff}

The purpose of the present section is to study the formfactor of the classical  action spectrum (\ref {eq:clff}).
We start by investigating alternative methods for averaging the formfactor, the need for which, and the difficulties
involved, were discussed at length in section (\ref {introduction}).  We shall study the function
${\cal K}_n(k)$ (\ref {eq:kcal}), and in particular test whether the action spectrum of the Baker map, satisfies
either of the equivalent relations
\begin{equation}
\label{sca4}
{\cal K}_n(k)\approx \frac{k}{2\pi n}K_{RMT}\left(\frac{2\pi n}{k}\right) \ ,\ {\rm or}\ ,\ \tau {\cal K}_n\left(\frac{2\pi
n}{\tau}\right)\approx K_{RMT}\left(\tau \right).
\end{equation}
 Since the Baker map is invariant under time reversal, the appropriate RMT expression is the one for the Circular
Orthogonal Ensemble (COE) $K_{\mathrm{COE}}(2 \tau)$, where the factor $2$ in the argument is due to the invariance
of the Baker map under the $\hat R$ symmetry. The explicit expression for $K_{\mathrm{COE}}(\tau)$ is given by
\begin{equation}
\label{sca6}
K_{\mathrm{COE}}(\tau)=\left\{ \begin{array}{lr}2\tau-\tau\ln(1\!+\!2\tau)&\mbox{  for
}\tau<1\vspace{2mm}\\ 2- \tau\ln\left(\frac{2\tau+1}{2\tau-1}\right)&\mbox{  for
}\tau> 1\end{array}\right. \ .
\end{equation}

\begin{figure}[h]
\centerline{\psfig{figure=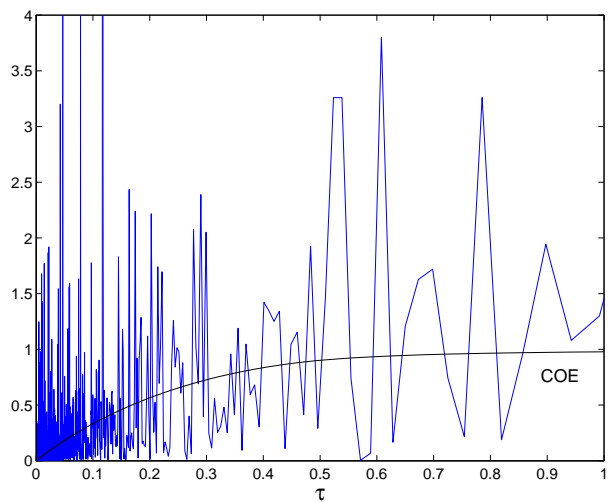,width=8cm}}
 \vspace{-3mm}
\caption{The function $\tau {\cal K}_n\left(\frac{2\pi
n}{\tau}\right) $  (n=15) (\ref {sca4}) and the  RMT prediction}
\label{fig:Ktaur1}
\end{figure}

Before applying any averaging, the function $\tau {\cal K}_n\left(\frac{2\pi n}{\tau}\right) $ displays strong
fluctuations which are shown in figure \ref{fig:Ktaur1}. for $n=15$ together with the expected COE result. The large
fluctuations make the comparison quite meaningless, and the figure is shown to emphasize the need of averaging.

\begin{figure}[h]
\centerline{\psfig{figure=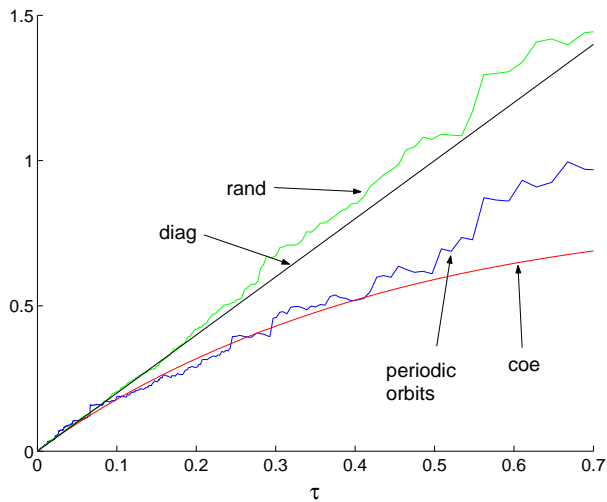,width=8cm}}
 \vspace{-3mm}
\caption{A running average (\ref {eq:running}) of  $  \tau {\cal
K}_n\left(\frac{2\pi n}{\tau}\right) $ (n=17)} \label{fig:Ktaucoe}
\end{figure}

Using the running average:
\begin{equation}
\label{eq:running}
\bar f(x) =\frac{1}{x}\int_0^x f(y){\rm d} y \ ,
\end{equation}
the large fluctuations are reduced, and the curve labelled
``periodic orbits" in figure \ref {fig:Ktaucoe}. is the running
average obtained by computing the formfactor from the set of
$n=17$-periodic orbits. The agreement with the corresponding COE
curve persists up to $\tau \approx 0.5$. The line marked ``diag"
is the curve obtained by assuming that the actions are not
correlated, and it agrees very well with the line marked as
``rand"  which was computed for the random set of actions (see
section (\ref {nndistribution})). The difference between the data
and the random curves is a clear indication of the presence of
correlations, whose similarity to the predicted COE result goes
beyond the leading ``diagonal" approximation.  The running average
procedure is not satisfactory, because  it is practical only for
low values of $n$. As $n$ increases, the number of periodic orbits
proliferates exponentially, and the fluctuations in the formfactor
grow as rapidly. Moreover, this method does not help to unravel
the dynamical origin of the correlations.

A systematic averaging procedure, which can be applied for any $n$, uses the concept of families
which was introduced in section (\ref{subsect:fam-codes}). The spectrum of actions of $n$-periodic orbits is
partitioned to families at the $r$'th level with labels ${\mathbf{P}}^{(r)}$ (see (\ref {eq:Pcode})).  As long as
$r$ is  sufficiently small such that  $2^{-r}$ is larger than the correlation length, only {\it intra} family
correlations are important. The formfactor can be approximated by an incoherent sum over the formfactors which
pertain separately to orbits within one family.
\begin{equation}
{\cal K}_n^{(r)}(k) = \frac {1}{ N_n}\  \sum_{ {\mathbf{P}}^{(r)} }\ \left | \sum_{a \in   {\mathbf{P}}^{(r)} } {\rm
e}^{i k s^{(n)}_a} \right |^2 \ .
\end{equation}
 Each family of actions is supported on an interval of size which decreases with $r$ as $2^{-r}$. Hence from the
requirement $k \Delta s > 2\pi$ we deduce
$N > 2^r$, and the range of values of $\tau$ which can be described by this method is bounded
from above by $\approx \frac{n}{2^r}$. Thus by increasing $r$, we gain a higher level of smoothing, but we lose
on the range of $\tau$ where this method can be used.

\begin{figure}[h]
\centerline{\psfig{figure=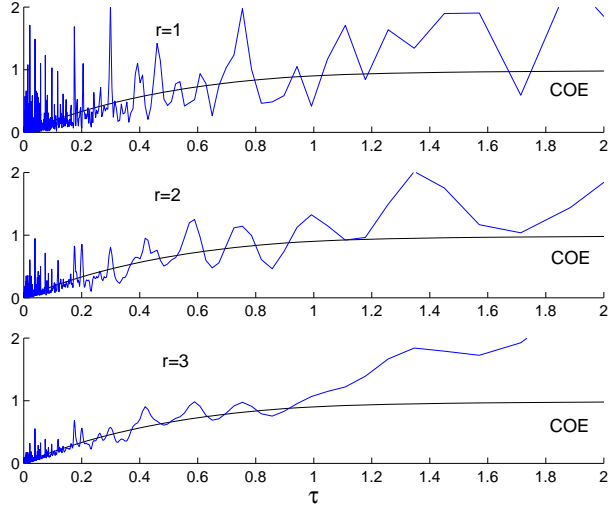,width=8cm}}
\vspace{-3mm}
\caption{${\mathbf{P}}^{(r)}$ family averaged $  \tau {\cal K}_n
\left(\frac{2\pi n}{\tau}\right)$ for $r=1,2,3 \ , \ n=15$}
\label{fig:Ktaucoef}
\end{figure}

\begin{figure}[h]
\centerline{\psfig{figure=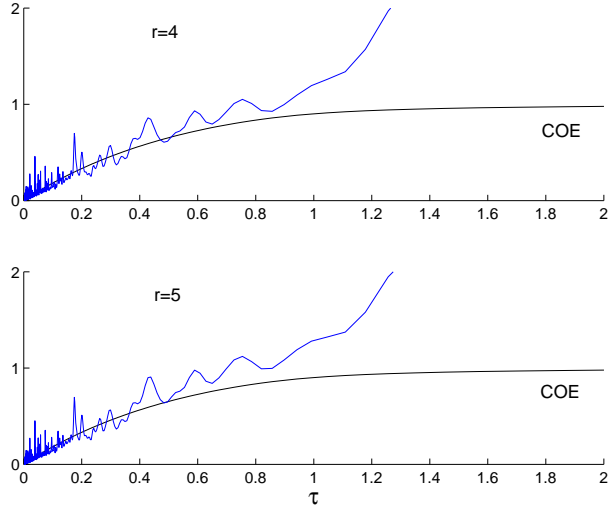,width=8cm}}
\vspace{-3mm}
\caption{${\mathbf{Q}}^{(r)}$ family averaged $  \tau {\cal
K}_n\left(\frac{2\pi n}{\tau}\right) $ for $r=4,5$ and $n=15$}
\label{fig:Ktaucoef45}
\end{figure}

    The resulting formfactors for $n=15$ and $r=1,2,3$ are shown in figure \ref {fig:Ktaucoef}. It shows that the
smoothing gets more effective as $r$ increases, without appreciable loss of correlations in the interval $0<\tau
<1 $. This is consistent with the expectation that the spectral correlation length is $\frac {1}{15}$, which
is smaller than the family separation of $\approx 2^{-r}$ for the values $r=1,2,3$. A closer comparison of
the $r=2$ and $r=3$ curves near $\tau=1$ shows that the  deviations from the RMT prediction starts
earlier for the $r=3$ data. To investigate this trend further, we use the
${\mathbf{Q}}^{(r)}$  partitioning (\ref {eq:Qcode}) which is more convenient
from the numerical point of view. It allows us to extend the level further, and the data for
$r=4,5$ is shown in figure \ref {fig:Ktaucoef45}. Since now the family range is smaller than $\frac{1}{n}$, the correlations
can be studied only on a lower range of $\tau$ value.  This, and similar other numerical investigations
provide strong evidence in support of the $\frac{1}{n}$ dependence of the correlation length, in agreement with the
semiclassical expectations.

 As was commented previously, the partition into families breaks pairs of $\hat R$ conjugate orbits. Hence, the
RMT expression which is used in figures  \ref {fig:Ktaucoef}. , \ref {fig:Ktaucoef45}. is  $K_{\mathrm{COE}}(\tau)$ and not
$K_{\mathrm{COE}}(2 \tau)$ which is used for the comparison with the full data set.

\begin{figure}[h]
\centerline{\psfig{figure=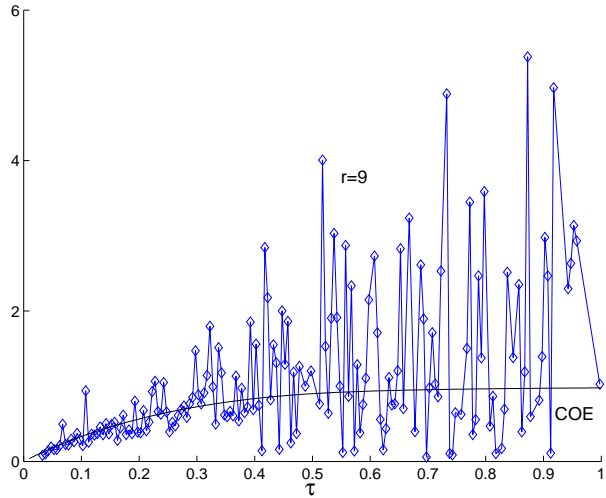,width=8cm}}
\vspace{-3mm}
\caption{ $ \tau {\cal K}_n\left(\frac{2\pi n}{\tau}\right) $,  from the Ising model with $r=9$, averaged over $17\le
n\le 22$.}
\label{fig:Ktaumod}
\end{figure}

 The Ising model approach (\ref {sec:Ising}) offers a complementary view of the formfactor. As was explained in
(\ref {sec:Ising}), the truncation at the $r$'th level does not allow us to distinguish between actions which are
within
$2^{-r}$, and counts them as if they are degenerate. Thus, the only way by which we can get spectral information on
the scale of
$\frac{1}{n}$ in this model is to use $r$ such that $\frac{1}{n} >> 2^{-r}$. The spectral correlations are now
studied as correlations {\it between} degeneracy groups, rather than {\it within} a group.
 Figure ~\ref{fig:Ktaumod}. shows the average of a few form factors, corresponding to values of $n$ from $n=17$ to
$n=22$, given by the model with $r=9$ which satisfies the criterion above. The average over $n$ reduces the
fluctuations of the form factor so it is possible to compare it directly to the $RMT$.

The fact that the action spectrum is resolved with an accuracy of $2^{-r}$ sets a lower limit to the values of
$\tau$ which are accessible by the model, which is $\tau_{min} \approx \frac{n}{2^r}$. This is why the low $\tau$
values in  figure  \ref{fig:Ktaumod}.  are absent.

\section{Discussion}
 The semiclassical predictions which were presented in section (\ref {introduction})
were checked in detail and were found to be well satisfied for the Baker map. They consist of the following main
points:
\begin{itemize}
\item The action spectrum show pair correlations which are universal and consistent with the predictions of the
semiclassical theory and RMT.
\item The  correlation length is of order $\frac {1}{n}$ which exceeds the mean spacing by a factor which grows
exponentially with $n$. The spectrum of actions is Poissonian  on smaller scales.
\item The correlations can be associated with families of periodic orbits which have a similar dynamical structure
which can be associated systematically with their symbolic codes.
\end{itemize}

 Various aspects of these points were confirmed for other systems \cite {argamandittes, dittesdoron, tanner,
cohenprimack,primack}. The results of \cite {cohenprimack,primack} are unique, because the phase
space of the mapping considered is four dimensional,  and the semiclassical theory predicts that the action
correlation range scales with  $n$ as $\frac{1}{n^2}$. This scaling was confirmed.

  Richter and Sieber \cite {SR2001} pointed out the pairs of periodic orbits which provide the correlations which
are necessary for the second order term in the small $\tau$ expansion of the formfactor. These orbits spend about
half of their time quite close to each other, and the rest of the time, they go along the time reversed part of
their partner's orbit. This way both orbits visit the same parts of phase space, but the order by which they do it
is different. Similar pairs were also proved to provide the answer in the case of quantum graphs \cite {Berko2002}.
The families which we identified here as storing the correlations between the actions are of the same nature, but of
a more general character. They do not rely exclusively on pairs conjugated partially by time reversal (as explained
above), but by the requirement that the correlated members spend the same amount of time in the same sub domains of
 phase space. This definition of families is also consistent with the work of \cite {cohenprimack}, where the
periodic orbits in a family follow the same segments of phase space which were related to each other by symmetries
other than time reversal.  Because of the strong link between phase space partition and symbolic codes, it
was possible to characterize the families by a common code. Unfortunately, even with this tool, we were not
able to derive the classical correlation function using classical arguments only. This remains an enigma
which should be addressed. At the same time, one should also be able to show that the spectrum of actions on
smaller scales is Poissonian. This is assumed, but not discussed in \cite {SR2001} and in \cite {Berko2002}.

In the present article we focussed our attention on discrete maps and their periodic orbits. However, most
system of interest are Hamiltonian flows, where the time is a continuous variable, and the quantum spectrum is on
the real line and not on the unit circle. This does not pose any essential problem, since several
quantization techniques make use of an auxiliary map to derive the quantum energies, and it was shown that
for chaotic systems, the spectral statistics of the energies and of the eigenphases of the auxiliary map
are the same in the semiclassical limit \cite {Bogomol,DoronUS,smil1}. A more direct approach was recently
introduced in \cite {BrunoUS}, where the  quantization of the hamiltonian flow is carried out in terms of
a quantum map which evolves the system along a sequence of equally spaced times $t_n\ =\ n\  \Delta t$. The
semiclassical expression for the spectral formfactor is  analogous to (\ref
{eq:duality}), and the periodic orbits have  periods which are integer multiples of $\Delta t$ and
their energies are restricted to a well defined energy interval.

 Integrable dynamics lead to Poissonian spectra  \cite {Berrytabor}. In the present context this implies that the
actions are  Poissonian too  \cite {cohenprimack}.  Recently it was shown in \cite {bogocor}
that in order to account for finer spectral correlations which are due to the spectrum being
pure point, finer correlations must exist, but this discussion exceeds the scope of the present
paper.

\section*{Acknowledgments}
%
 We would like to acknowledge discussions with and comments from Herve Kunz, Gergely Palla and Jean-Luc Helfer. The
work was supported by the Minerva center for Physics of Complex Systems and by grants from the Israel Science
Foundation and the Minerva Foundation.




\begin{thebibliography}{99}
\bibitem {bgs}  O. Bohigas \emph{Random matrix theories and chaotic dynamics},  Les
Houches, Session LII, Elsevier Science Publishers B.V. (1989)
\bibitem{argamandittes} N. Argaman, F.M. Dittes, E. Doron, J.P. Keating, A.Y. Kitaev, M. Sieber and U. Smilansky \emph{PRL}
 \textbf{71}, 26 ,4326-4329 (1993)
\bibitem{dittesdoron}  F.M. Dittes, E. Doron and U. Smilansky \emph{Phys.Rev.E} \textbf{49}, 2, 963-966 (1994)
\bibitem {aurich} R. Aurich and M. Sieber,  {\em J. Phys. A} {\bf  27 } (1994) 1967--1979.
\bibitem{smil1}  U. Smilansky \emph{Semiclassical Quantization of Chaotic Billiards - a Scattering Approach},
 Les Houches,
Session LXI, Elsevier Science Publishers B.V. (1995)
\bibitem{tanner}  G. Tanner \emph{J.Phys.A} \textbf{32} , 5071-5085 (1999)
\bibitem{cohenprimack}  D. Cohen, H. Primack and U. Smilansky, \emph{Ann.Physics} \textbf{264}, 108-170 (1998)
\bibitem{sano} M.M. Sano \emph{Chaos} \textbf{10}, 1, 195-210 (2000)
\bibitem{primack}  H. Primack and U. Smilansky \emph{Phys.Reports} \textbf{327} , 1-2  (2000)
\bibitem{harayama}  T. Harayama and A. Shudo \emph{J.Phys.A} \textbf{25}, 4595 (1992)
\bibitem{SR2001}  M. Sieber and K. Richter {\it Phys. Scr.} {\bf 90} 128, (2001)
\bibitem {Braun2002}  P. A. Braun, F. Haake and S. Heusler \emph{J. math Phys} {\bf 35} 1381,(2002).
\bibitem{Berko2002} G. Berkolaiko, H. Schanz and R. Withney, {\emph Phys. Rev. Letters} {\bf 82}, 104101, (2002).
\bibitem{smil2}  U. Smilansky \emph{Semiclassical Quantization of maps and spectral correlations}
 Proc. of the Nato Advanced Study  Institute  ``Supersymmetry and Trace Formulae", Editor I. Lerner Cambridge, (1997)
 (in press)  (2002).
\bibitem{gold} H. Goldstein \emph{Classical Mechanics} Addison-Wesley, Reading, MA, (1957)
\bibitem{saraceno} M. Saraceno \emph{Ann.Physics} \textbf{199} , 37-60  (1990)
\bibitem{balazsvoros} N.L. Balazs and A. Voros \emph{Ann.Physics} \textbf{190} , 1-31  (1989)
\bibitem{saracenovoros} M. Saraceno and A. Voros \emph{Physica D} \textbf{79} , 206-268 (1994)
\bibitem{dittesdoron2}  F.M. Dittes, E. Doron (private communication)
\bibitem{berry} M.V. Berry \emph{Proc. R. Soc. Lond. A} \textbf{400}, 229-251 (1985)
\bibitem{accuracy} H. Primack and U. Smilansky  On the accuracy of the semiclassical trace formula.
J. Phys. A: Math. Gen. {\bf 31},6253-6277,  (1998).
\bibitem{hannayozorio} J.H. Hannay and M. Ozorio De Almeida \emph{J. Phys. A} \textbf{17}, 3429-3440 (1984)
\bibitem {Sinailect} {\em Geometric Aspects of Functional Analysis} Lecture Notes
in Math. 1469,  Springer, Berlin 41-59 (1990)
\bibitem{ozoriosareceno} M. Ozorio De Almeida and M. Saraceno \emph{Ann.Physics} \textbf{210} , 1-15  (1991)
\bibitem {Bogomol} E.B. Bogomolny   {\em Nonlinearity} {\bf 5} (1992) 805.
\bibitem {DoronUS}  E. Doron and U. Smilansky.  {\em Nonlinearity} {\bf 5} (1992) 1055.
\bibitem {BrunoUS}  B. Eckhardt and U. Smilansky   {\em Foundations of Physics}, {\bf 31} 543-556 (2001).
\bibitem {Berrytabor} M. V. Berry and M. Tabor {\em Proc. Roy. Soc.} {\bf A349} 101-123 (1976)
\bibitem {bogocor}   E. Bogomolny, Action correlations in integrable systems, preprint
(1999) (chao-dyn/9910036).
\end{thebibliography}
\end{document}